RESEARCH ARTICLE

# Irreversible Collective Migration of Cyanobacteria in Eutrophic Conditions

Julien Dervaux[1]*, Annick Mejean[1], Philippe Brunet[2]

**1** Laboratoire Interdisciplinaire des Energies de Demain, Université Denis Diderot, Paris, France,
**2** Laboratoire Matière et Systèmes Complexes, Université Denis Diderot, Paris, France

* dervauxjulien@wanadoo.fr

## Abstract

In response to natural or anthropocentric pollutions coupled to global climate changes, microorganisms from aquatic environments can suddenly accumulate on water surface. These dense suspensions, known as blooms, are harmful to ecosystems and significantly degrade the quality of water resources. In order to determine the physico-chemical parameters involved in their formation and quantitatively predict their appearance, we successfully reproduced irreversible cyanobacterial blooms in vitro. By combining chemical, biochemical and hydrodynamic evidences, we identify a mechanism, unrelated to the presence of internal gas vesicles, allowing the sudden collective upward migration in test tubes of several cyanobacterial strains (*Microcystis aeruginosa* PCC 7005, *Microcystis aeruginosa* PCC 7806 and *Synechocystis sp*. PCC 6803). The final state consists in a foamy layer of biomass at the air-liquid interface, in which micro-organisms remain alive for weeks, the medium lying below being almost completely depleted of cyanobacteria. These "laboratory blooms" start with the aggregation of cells at high ionic force in cyanobacterial strains that produce anionic extracellular polymeric substances (EPS). Under appropriate conditions of nutrients and light intensity, the high photosynthetic activity within cell clusters leads the dissolved oxygen (DO) to supersaturate and to nucleate into bubbles. Trapped within the EPS, these bubbles grow until their buoyancy pulls the biomass towards the free surface. By investigating a wide range of spatially homogeneous environmental conditions (illumination, salinity, cell and nutrient concentration) we identify species-dependent thresholds and timescales for bloom formation. We conclude on the relevance of such results for cyanobacterial bloom formation in the environment and we propose an efficient method for biomass harvesting in bioreactors.

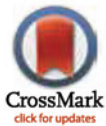

**Data Availability Statement:** All relevant data are within the paper and its Supporting Information files.

**Funding:** This project is funded by the program Emergence(s) of the City of Paris. The funders had no role in study design, data collection and analysis, decision to publish, or preparation of the manuscript.

**Competing Interests:** The authors have declared that no competing interests exist.

## Introduction

Within the last 60 years, in part following the development of modern agriculture and its counterpart in the massive use of soil fertilizers, seasonal blooms of aquatic photosynthetic microorganisms have become commonplace in quiet rivers, lakes, fishponds and even the open ocean [1–3]. During a bloom, the phytoplankton density near the surface of the water column increases by several orders of magnitude within timescale of days from less than $10^2 - 10^3$ cell/ml up to $10^5 - 10^9$ cell/ml and sometimes culminates into a thick superficial foam [4]. These episodes of massive biomass accumulation involve various species like micro-algae, cyanobacteria or protists [5] and multiple evidences [2, 6–12] suggest that the ability to bloom provides several competitive advantages over non-blooming organisms (access to light and $CO_2$, avoidance of grazing and ability to counter light mixing). Still, blooms are harmful upon various aspects: they shadow sunlight to plants and other micro-organisms, some species release powerful toxins that are harmful to humans, mammals as well as various marine species [13, 14] and biomass decay ultimately leads to oxygen rarefaction or even depletion (anoxia) that can stress or asphyxiate fishes, shellfishes or invertebrates [15].

From numerous environmental monitoring, it is now well established that blooms are correlated with a combination of intense sunlight, high water temperatures, eutrophic conditions and gentle stream flows [16, 17]. They are often observed during late spring or summer, near agricultural regions where nitrate or phosphate-based fertilizers are carried by water runoffs and give micro-organisms a sudden growth boost. However, because the characteristic timescale of blooms (about 1 day) is not compatible with cell proliferation alone (cell division times range from 10h in optimal laboratory conditions to a few days for most bloom-associated species [4]), the sudden increase of the biomass concentration within the epilimnion (the superficial water) is likely to result from an interplay between long-term multiplication of cells and the collective migration of the biomass. In many situations, this migration culminates in the irreversible formation of a thick scum at the surface of water. While it has been clearly established that buoyancy regulation through the synthesis of gas vesicles and carbohydrate ballast is a key factor involved in the massive migration of the biomass [18–22], several non-vacuolate cyanobacterial species from various genera, such as *Synechococcus* spp., *Aphanothece* spp, as well as other species [23], are nonetheless able to form blooms. This suggests that one or several other mechanisms are likely to contribute to the process of bloom formation. In addition, it is still unclear how a dense suspension of micro-organisms, as formed by the collective upward migration of the biomass, can further evolve into the robust compact scum which is responsible for the most harmful effects associated with blooms.

Therefore, the dynamics of the migration process should be precisely recorded in order to better understand the blooming mechanism. In the seek for accurate predictions of their formation, laboratory studies can provide a valuable complement to field observations. However, because the spatiotemporal monitoring of micro-organisms concentration during bloom formation is challenging, only few laboratory experiments have been performed so far [24, 25] and the accurate measurement of local density of micro-organisms remains to be achieved.

The approach followed in the present paper is to investigate quantitatively to what extent such thick scum can irreversibly form out of a collective migration and agglomeration, from well-controlled initial conditions compatible with environmental ones. Therefore, we mainly focus on the last stages of this scum formation. By carrying out systematic milliliter-scale experiments in test tubes, we investigate a range of environmental conditions (initial concentration of microorganisms, light intensity, salinity, nutrients chemical composition and concentration) and we identify those leading to an irreversible migration toward the surface in cultures of the cyanobacteria *Microcystis aeruginosa* PCC 7005. Indeed, members of the *Microcystis* genus are frequently detected in blooms appearing in rivers and freshwater lakes [5, 26, 27]. We also evidence the same behavior with the toxic strain *Microcystis aeruginosa* PCC 7806 that produces the hepatotoxin microcystin [28, 29] and with the unrelated non-vacuolate specie *Synechocystis sp.* PCC 6803. When blooming occurs, initially homogeneous liquid cultures phase-separate into dense foamy layers floating above bacteria-free liquid medium (Fig. 1-A) and the timescale of this migration was recorded for each experiment. By combining this extensive dataset with *in situ* oxygen measurements, EPS concentration and surface charge quantification, rheology experiments and microscopic observations, we rule out the influence of gas vesicles on this process and evidence the coupled roles of photosynthetically-produced oxygen accumulation and anionic EPS-mediated cell aggregation following salt addition on blooming. The mechanism of collective upward biomass migration identified here relies on the trapping of oxygen bubbles within cell clusters that provide a buoyant lift to the biomass and can occur within timescales as short as 30 minutes.

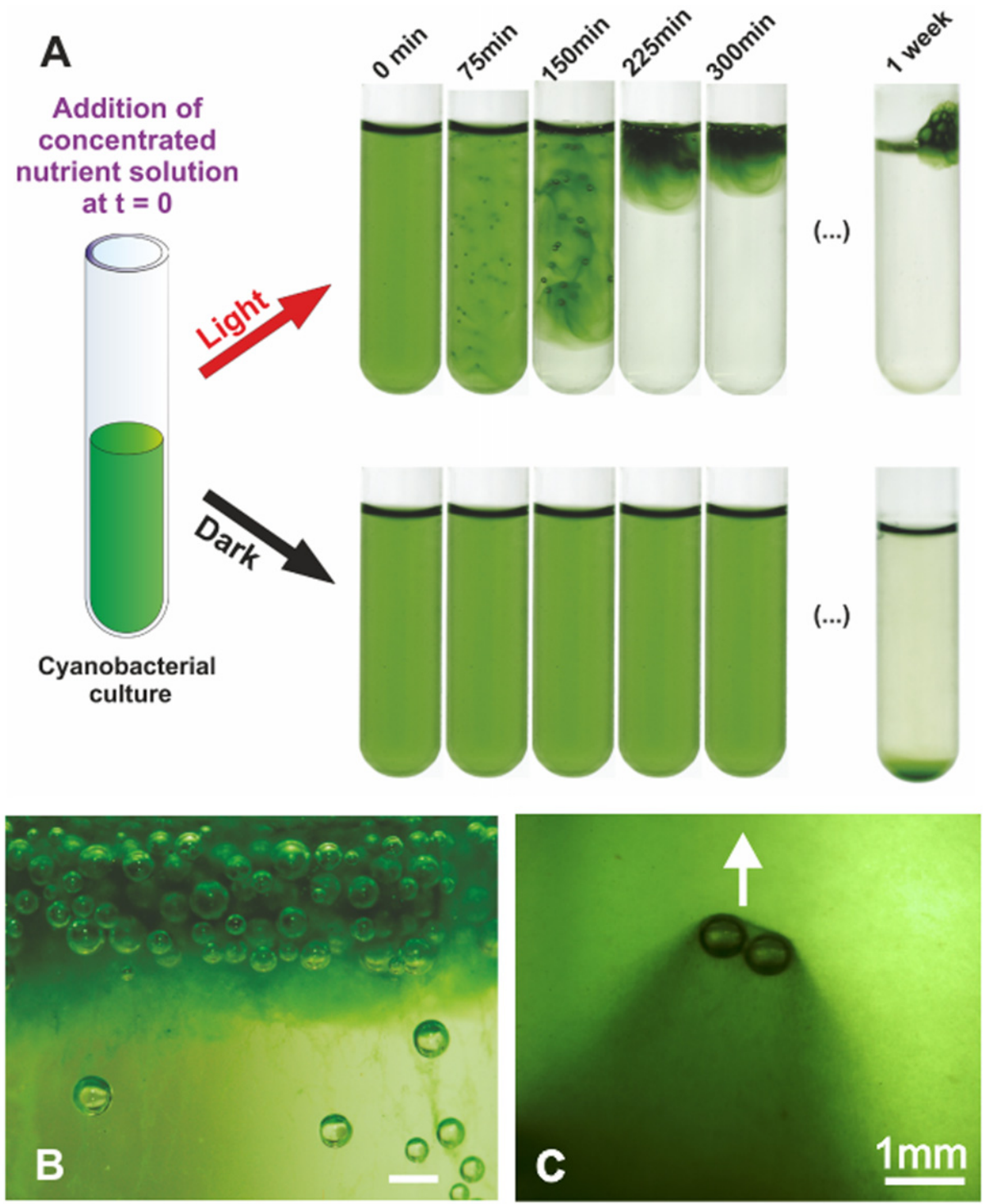

**Fig 1. *M. aeruginosa* bloom in a test tube.** A: Upward migration of the whole biomass following the addition of nutrients, inducing the irreversible formation of a scum stable over weeks, with *M. aeruginosa* PCC 7005. The initial $OD_{580}$ of the cyanobacterial culture is 0.66. This blooming is not observed in darkness. See text for more details. B: Close up view of the topmost foamy layer, the scale bar is 1mm C: Trail of dense biomass around rising bubbles. The arrow indicates the direction of bubble migration.

## Materials and Methods

### Strains and culture conditions

Strains used in this study are *Microcystis aeruginosa* PCC 7005, *Microcystis aeruginosa* PCC 7806, *Synechocystis sp.* PCC 6803. All strains were grown in BG11 (1X) medium, with trace metal solution, under a 14h/10h light cycle ($6 \ \mu E.m^{-2}.s^{-1}$) at 25°C. The composition of the BG11 medium (1X) is as follow: $CaCl_2.2H_2O$ (36.7mg/L), citric acid (5.6 mg/L), $K_2HPO_4$ (31.4mg/L), $Na_2Mg.EDTA$ (1mg/L), $C_6H_8FeNO_7$ (6mg/L), $MgSO_4$ (36mg/L), $Na_2CO_3$ (20mg/L), $NaNO_3$ (1500mg/L). The trace metal solution is composed of: $H_3BO_3$ (2860 mg/L), $MnCl_2—4H_2O$ (1810 mg/L), $ZnSO_4—7 \ H_2O$ (222 mg/L), $Na_2MoO_4—2 \ H_2O$ (390 mg/L), $CuSO_4- 5 \ H_2O$ (79 mg/L), $Co(NO_3)_2$-6H2O (49 mg/L).

### Biomass measurement

Optical density measurements were performed at 580 nm [30, 31] on a Novaspec II spectro-photometer (Amersham Pharmacia Biotech Inc, UK) using a cuvette with a 1cm light path. Cell counting was performed with a Malassez cell on an inverted microscope and gave a concentration of $2 \cdot 10^7$ cell/ml at an optical density $OD_{580} = 1$. Using an average cell diameter of $5 \mu m$, this corresponds to a volume fraction of $\sim 10^{-3}$.

### Sample preparation for bloom formation

All experiments were performed at 25°C. For the determination of the salt thresholds, aliquots from culture in the late exponential phase ($OD_{580} \sim 2$) were diluted to a final concentration of $OD_{580} = 0.35$. The influence of the gas vesicles on blooming—using ultrasonic waves to make them collapsing—was tested on aliquots from cultures in middle and late exponential phases ($OD_{580}$ of 1 and 2) which were diluted to a final concentration of $OD_{580} = 0.35$. Sonication was conducted in a water bath sonicator at 25kHz for 5 minutes. In all experiment the pH of the bacterial suspension was comprised between 7.0 and 8.0. The light and nutrient concentrations for each experiments are specified in the main text. In all experiments, the light source (a large LED pannel) was located on the side of the test tubes and provided a highly homogeneous lightning.

### Oxygen measurements

Oxygen measurements presented in Fig 2 were performed using a Multi 3430 sensor equipped with an FDO925 optode (WTW, Germany). Because of the size of the probe, its width being comparable the tubes diameter, the continuous oxygen measurements were conducted in 10ml glass bottle containing 6 ml of cultures. The experiments were performed in triplicate and the error bars are the standard deviations between the three runs.

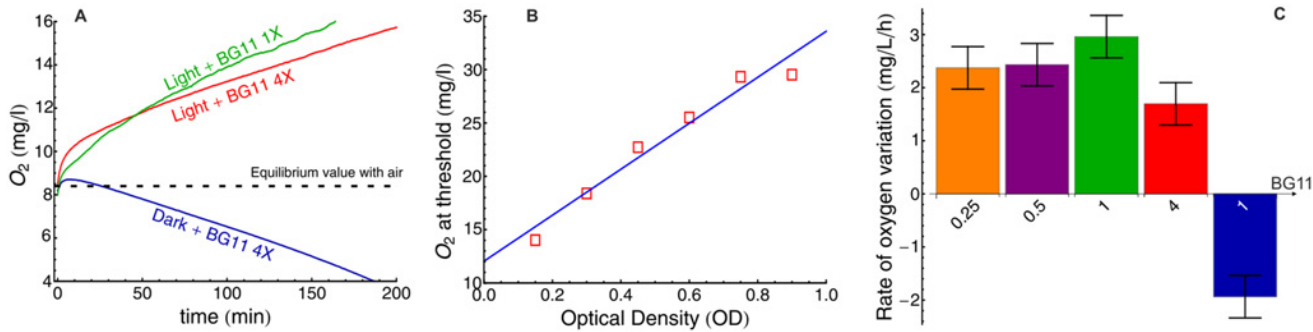

**Fig 2. Oxygen bubbles lift the biomass during bloom formation.** A: $O_2$ production following BG11 addition to cultures of *M. aeruginosa* PCC7005 (final $OD_{580} = 0.35$). B: Dissolved Oxygen (DO) concentration at the threshold of bloom formation as a function of initial optical density, evidencing that the lifting force is mainly due to these bubbles. The threshold was defined as the time at which the veil at the bottom of the biomass is located 5 mm above the bottom of the test tube. When the threshold was reached, the sample was carefully diluted in BG11 (1X) and placed in the dark to measure the DO concentration. C: Rate of $O_2$ production ($OD_{580} = 0.35$) for various BG11 concentration (values at the foot of each bar) in the light or in the dark (right bar in blue).

## Imaging techniques

Images were acquired with a Nikon D90 equipped with a Makro-Planar 2/50 ZF Zeiss objective except panel B of Fig. 1 (taken with a Tokina Macro 100 F2.8 D equipped with an extension tube) and panel C of Fig. 1 (taken with a Navitar objective). Microscopic images in panels A and B of Fig. 3 were acquired with a Olympus IX3 inverted microscope coupled to a Qimaging Optimos sCMOS camera. Fig. 3-D was acquired with a Leica MZ16 microscope.

## Rheology measurements

Rheology measurements were performed on a controlled shear stress rheometer (Anton-Paar Physica MCR500) in a cone and plate geometry with a 52 $\mu m$ gap. The samples were kept at 23.00°C during data acquisition. For the viscosity evolution experiments, rheology measurements were made on samples prepared as follow: 50$\mu l$ of concentrated BG11 solution (100X) were added to 1 ml of a sample from a culture in the late exponentially phase ($OD_{580} = 2.4$) in an Eppendorf tube and inverted upside-down three times. The tubes were then placed under a 13 $\mu E.m^{-2}.s^{-1}$ light. At intervals of 5 minutes, the content of the tubes was gently poured onto the rheometer plate and the viscosity was measured at a shear rate of $3 \cdot 10^{-2} s^{-1}$. The measurements were performed three time and the error bar is the standard deviation between the three runs.

## EPS extraction and quantification

EPS were extracted mostly like in [31, 32]. A sample of 30 ml of a stationary phase culture ($OD_{580} \sim 4$) was split in two volumes and blooming was triggered by adding 1.5ml of concentrated BG11 solution (100X) in one of the samples. Both samples were then placed under light (14 $\mu E.m^{-2}.s^{-1}$) for 90 minutes. The samples were then placed for 90s in a water bath sonicator at intermediate power (25kHz) to separate cells from loosely bound EPS. The samples were then centrifuged at 10000$g$ for 30 minutes at room temperature. The supernatant was separated from the cells and the centrifugation step was repeated twice. Following centrifugation, the cell pellets were resuspended in BG11 medium (1X) and the supernatant of the three runs were added. Then, the supernatants were filtered (using a paper filter of typical pore size 0.45 $\mu m$) and stored overnight at -20°C with two volumes of chilled ethanol to precipitate the EPS. The two tubes of precipitated EPS were then washed with 95% ethanol, dried under air and

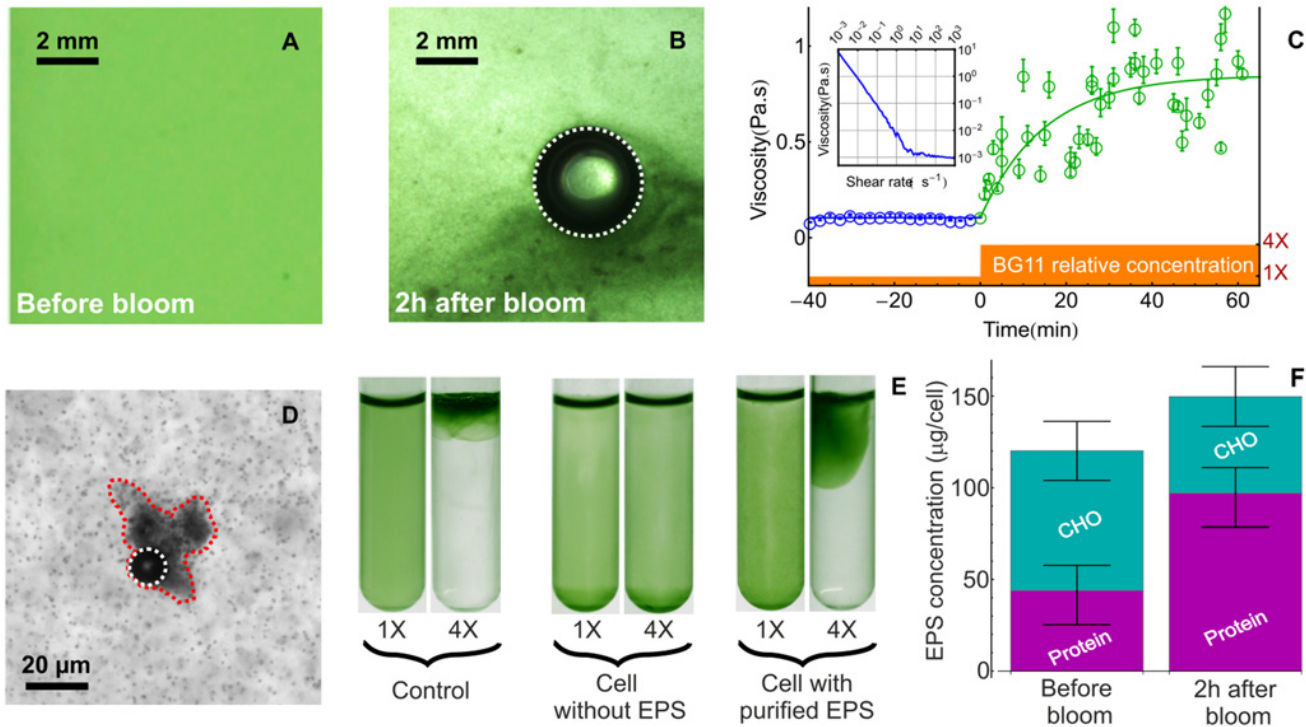

**Fig 3. Aggregation of the biomass during a bloom.** A-B: Microscopic details of the microstructure before (A) and following (B) BG11 addition (final relative concentration of 4X) to a cyanobacterial culture of *M. aeruginosa* (final $OD_{580}$ = 0.5). C: Time evolution of the viscosity of a cyanobacterial culture before (black circles) and after (green circles) nutrient addition. We first measured the viscosity of a concentrated culture ($OD_{580} \sim 2.4$) as a function of the shear rate (see Inset). The bacterial suspension exhibits a shear-thinning behavior in the range $10^{-3}$ to 1 s$^{-1}$ where the viscosity falls by 3 orders of magnitude. The apparent divergence of viscosity at low shear is the signature of a yield stress behavior and this can be ascribed to the interactions between EPS, bacterial cells and ions. After a crossover regime (1 to 10 s$^{-1}$) the viscosity becomes independent of the shear rate in the range 10 to $10^3$ s$^{-1}$ and saturate at 1.1±0.1 mPa.s. In this high-shear limit, the viscosity is very close to that of water as expected for a dilute (volume fraction around $2 \cdot 10^{-3}$) suspension of weakly motile bacteria. D: A bubble trapped in a cluster of cells (surrounded by a red dashed line) at high magnification. The contours of growing bubbles in B and D have been highlighted by white dashed lines. E: Cells were separated from their EPS using sonication and centrifugation. Without EPS, "naked" *M. aeruginosa* cells failed to form a bloom. When the "naked" cells were supplemented with the purified EPS, the ability to form a bloom was restored. Pictures were taken 180 minutes following BG11 addition. In the experiment with naked cells supplemented with purified EPS, the threshold for bloom formation remained identical to that of untreated (i.e with their EPS). F: Chemical composition of the EPS before and after blooming.

doi:10.1371/journal.pone.0120906.g003

resuspended in 500$\mu$L of ultra-pure water (UPW) each. The protein and carbohydrate contents of the EPS were quantified respectively with a Bradford assay using BSA as a reference and with a phenolsulfuric assay using glucose as a reference. Zeta potentials of purified EPS were measured with a zetasizer nano ZS (Malvern, UK). We found a zeta potential of −23±6mV that was neutralized upon BG11 addition. Experiments were performed in triplicate and error bars are standard deviations between the three runs.

## Results

### A sudden increase in salts concentration triggers blooming

Cultures of the non-toxic cyanobacterial strain *M. aeruginosa* PCC 7005 were grown in standard mineral medium (BG11+Trace Metals) until they reached an optical density at 580nm ($OD_{580}$) of 2 (i.e. $\sim 4.10^7$ cell/ml). Aliquots of the culture were then re-suspended in concentrated BG11 medium ($\sim$ 4 times the concentration of the initial culture) to a final $OD_{580}$ of 0.66 in 5ml test tubes. The tubes were placed either in complete darkness or under homogeneous lighting (intensity 13 $\mu$E.m$^{-2}$.s$^{-1}$) and imaged using a digital camera during 6 hours. Fig 1 illustrates the typical time evolution of the system. While the culture kept in darkness remained homogeneous for the duration of the experiment, we observed a massive migration of the biomass towards the free surface in the sample exposed to light.

This migration started with the nucleation of gas bubbles inside the culture tube which become visible roughly 60 minutes after the beginning of the experiment. Gentle shaking of the tubes revealed that bubbles were not stuck to the walls of the tubes but rather localized inside the bulk of the liquid, suggesting that they were somehow trapped by the biomass. These bubbles kept growing until they reached sizes of order of a millimeter (after roughly 150 minutes), at which point they started to rise, pulling the biomass upward with them. Trails of dense green biomass could be observed around and below rising bubbles (Fig 1-C). In addition, a sharp boundary delimited the bottom of the rising biomass, below which the culture liquid was transparent (Fig 1-A). The system stopped evolving after 300 minutes. We measured the optical density of the lower fraction after 300 minutes and found it to be 0.03 such that, in the final state, more than 95% of the entire biomass was located at the free surface. In a control tube (no BG11 added), the OD$_{580}$ was measured at 0.73 after 300 min so that the relative increase in biomass that occurred during the experiments was 11%. Because no bloom is formed when bacteria are resuspended in dilute nutrient solution, we then asked what was the minimum BG11 concentration necessary to induce the migration of the biomass. At an OD$_{580}$ of 0.35 and under $\sim$ 14 $\mu$E.m$^{-2}$.s$^{-1}$, blooming occurred at a relative BG11 concentration of $\sim$ 4X and above (S1 Fig). Remarkably, the cyanobacteria did not re-colonize the remaining part of the tube and stayed localized in the bacterial foam formed at the surface (Fig 1-B), even after one week. In the sample placed in darkness, sedimentation occurred but on a much longer timescale (one week). This slow sedimentation is consistent with the absence of gas vesicle *de novo* synthesis after a prolonged exposure to darkness [18, 19, 33].

## Trapped oxygen bubbles lift the biomass

Because *M. aeruginosa* has the ability to regulate its buoyancy using internal gas vesicles, we first tested whether this mechanism was contributing to the irreversible migration described above by collapsing the vesicles by sonication (see Materials section and S2 Fig). In a control tube (no concentrated BG11 added), sonicated cyanobacterial cells had completely sedimented after 300 minutes, therefore ensuring that i) the vesicles were indeed collapsed by the sonication treatment and ii) the cyanobacteria could not synthesize enough new vesicles to counter sedimentation over this timescale, in agreement with previous studies [18]. In another sample sonicated identically, but with BG11 added, blooming occurred nonetheless and the biomass was entirely concentrated at the free surface after 300 minutes. Furthermore, we compared the biomass migration with and without previous sonication, for samples with BG11 added, and no differences were observed between the two samples. From this assay, we concluded that the blooming mechanism identified here does not rely on the presence of gas vesicles.

We then investigated the origin of the gas observed during bloom formation. We first noted that no bloom or bubble were formed in the dark or when photosynthesis was inhibited by the addition of 5$\mu$M of DCMU (3-(3,4-dichlorophenyl)-1,1-dimethylurea). We therefore hypothesized that the gas was oxygen, a byproduct of the photosynthetic activity. We therefore used an oxygen sensor to measure the time evolution of the dissolved oxygen concentration in a cyanobacterial culture following the addition of the concentrated nutrient solution, both in the dark and under illumination (13 $\mu$E.m$^{-2}$.s$^{-1}$). The data presented in Fig 2-A shows that DO concentration was initially around 8.3 mg/L (the saturation value of DO for a BG11 solution in equilibrium with air) and increased immediately and rapidly following the resuspension of the bacteria in fresh concentrated medium. In the dark, DO concentration stopped increasing within 5 minutes of the light extinction and then decayed below the equilibrium value of 8.3 mg/L and until oxygen was completely depleted, thereby indicating active aerobic respiration. On the other hand, when the culture was exposed to light, the DO concentration increased linearly, at a rate of $\sim$ 2mg/L/h, until it reached the detection limit of the DO probe (16 mg/L), roughly 200 min after the beginning of the experiment.

Since DO was present at concentrations well above the saturation value, we concluded that the nucleation of $O_2$ bubbles that could lift the biomass was thermodynamically favored. To check whether $O_2$ bubbles were indeed responsible for the biomass migration, we thus investigated the relationship between DO concentration and total biomass at the threshold of bloom formation. If $O_2$ is responsible for the vertical migration of the cyanobacteria, we expect its concentration to be proportional to the total weight of this biomass at the onset of bloom formation. As seen in Fig. 2-B, there is indeed a very good correlation (P-value $< 10^{-3}$) between these quantities. Furthermore, when a sample was first placed in darkness for 12 hours following the addition of the concentrated BG11 solution to bring the DO to 0 mg/L, it took a much longer time ($\sim$ 8 hrs) before bloom formation occurred, further strengthening the evidence that $O_2$ bubbles lift the biomass. In addition, the DO concentration at threshold was similar to that obtained without keeping the sample in darkness at first. We then questioned whether the sudden increase in nutrient was leading to an increase of the oxygen production rate on the time scale of the experiment and/or was only responsible for the trapping of the $O_2$ produced by the photosynthetic activity of the bacteria. We therefore measured the rate of DO variation at various BG11 concentrations. Within the range of BG11 concentrations we investigated (0.25X to 4X), no significant variations in the DO production rate was observed (average value 2.4±0.4 mg/L/h for a culture at $OD_{580}$ = 0.35). In darkness, the rate of $O_2$ uptake was measured at 1.9±0.4 mg/L/h.

## Aggregation of the biomass leads to bubble trapping in the bulk

Since the sudden addition of nutrient had no significant effect on oxygen production on the timescale of the experiment, we were led to the conclusion that the increase in BG11 concentration was essentially responsible for providing bubble nucleation site within the biomass. A microscopic inspection of the cyanobacterial culture before (Fig. 3-A) and after (Fig. 3-B and D) the addition of the concentrated nutrients revealed that oxygen bubbles were indeed trapped in large aggregates of cell. Furthermore, while cells within clusters were physically linked, as could be seen by gently moving the bubble, they were not necessarily in contact, suggesting that cell to cell binding must be mediated by extracellular components such as EPS. In order to quantify the extent and dynamics of this aggregation, we monitored the rheological properties of the cyanobacterial culture over time as a proxy for the evolution of the biomass microstructure. The data shown in Fig. 3-C indicate that the low-shear viscosity was constant around $\sim$ 0.1Pa. s before nutrient addition. Following the sudden addition of BG11, the viscosity increased roughly by an order of magnitude. Fitting the data with a simple exponential yielded a characteristic timescale of 15min for the aggregation dynamic. Given the typical metabolic timescales of *M. aeruginosa* (doubling time $\sim$ 50 hours), this timescale is not compatible with the production of significant amounts of extracellular materials. To further establish that irreversible blooming depends on the presence of extracellular components present before BG11 addition, we first separated the cells from the EPS. As shown in Fig. 3-E, cells without EPS failed to bloom following BG11 addition. However, the ability to bloom was restored by supplementing the cell with purified EPS. The carbohydrate and protein content of the EPS was analyzed before and 2 hours after BG11 addition. As seen in Fig. 3-F the total EPS fraction is approximately constant at $\sim 130 \pm 30\text{pg.cell}^{-1}$. Although the moderate increase of the EPS ($\sim 25\%$) is within the error bar of the measurement, it could also be partly caused by a regulation of the transcription processes and the production of the EPS by metallic cations, as observed in other cyanobacterial species [34].

## Anionic EPS-cations interactions trigger biomass flocculation

In order to further pinpoint the mechanisms leading to the aggregation and blooming of the biomass, we individually tested the effect of the various components of the BG11 nutritive medium. It clearly appeared that sodium nitrate could induce bloom formation although at a slightly higher threshold than when $NaNO_3$ was dispensed with the other components of the BG11 medium. This reveals a synergistic effect between the various salts present in the BG11 medium. To discriminate between sodium and nitrate, we then assayed bloom formation following the addition of various other salts. The data presented in Table 1 show that $Na^+$, as well as other cations, could trigger the aggregation of the biomass. When the cation chelator EDTA was added to the culture with the salts, bloom formation was suppressed. Furthermore, electrophoretic mobility measurements performed on purified EPS indicated that they carry a negative charge, in agreement with other studies that have reported the presence of acidic sugars among cyanobacterial exopolysaccharides [35–37]. From these data, we concluded that aggregation occurred as a result of the flocculation between anionic EPS and nutrient-associated cations. Additionally, we found that the critical concentration for bloom formation was cation-dependent. Classifying the cation in order of flocculating power, we found: $Mg^{2+} \sim Ca^{2+} > K^+ \sim Na^+$. This is in qualitative agreement with the Schulze-Hardy rule stating that flocculation thresholds decrease with ion valency [38, 39].

## Irreversible blooming occurs in several cyanobacterial strains

Because the dual ability to synthesize anionic EPS and produce oxygen is not restricted to *M. aeruginosa* PCC 7005, we tested various other cyanobacterial strains and species. Bloom was successfully induced in the toxic stain *M. aeruginosa* PCC 7806, with the same salts used to trigger bloom formation in *M. aeruginosa* PCC 7005, although at much higher thresholds (e.g. a relative BG11 concentration of 14X against 4X for the strain PCC 7005). We also successfully induced bloom formation in the unrelated, non-vacuolate specie *Synechocystis sp.* PCC 6803, which further strengthened the evidences presented above that the migratory process studied here does not rely on the presence of gas vesicles. In this latter case however, collective migration only occurred upon addition of $Ca^{2+}$ or $Mg^{2+}$ but no blooms were observed when samples of PCC 6803 were supplemented with either BG11 (up to a relative concentration of 50X), $NaNO_3$, NaCl, KCl (up to concentrations of 1M). This indicate that divalent cations, but not

**Table 1. Salt and species-dependent concentration thresholds (in mM, except for the BG11 medium) above which irreversible blooms form.**

|  | BG11 | NaNO₃ | NaCl | KCl | CaCl₂ | MgCl₂ |
|---|---|---|---|---|---|---|
| *Microcystis aeruginosa* PCC 7005 | 4 ± 1 | 60 ± 5 | 60 ± 5 | 60 ± 5 | 20 ± 5 | 20 ± 5 |
| *Microcystis aeruginosa* PCC 7806 | 14 ± 1 | 350 ± 50 | 350 ± 50 | 400 ± 50 | 65 ± 5 | 55 ± 5 |
| *Synechocystis sp.* PCC 6803 | - | - | - | - | 5 ± 1 | 45 ± 1 |

Reference concentration of 1 is that of BG11 specified in Material and Methods. The reported concentrations of BG11 are the final relative values while the concentrations of all other salt are the amount added to the samples and are expressed in mM. For comparison, the BG11 at 1X contains 17.65 mM $NaNO_3$, $\sim 18$ mM $Na^+$, $\sim 0.2$ mM $K^+$, $\sim 0.25$ mM $Ca^{2+}$ and $\sim 0.3$ mM $Mg^{2+}$ monovalent ones, can reduce the electrostatic repulsion between PCC 6803 cells and lead to biomass aggregation and subsequent migration. Interestingly, when divalent cations were added above the threshold for bloom formation together with monovalent cations, no blooms occurred. This reveals that metallic cations compete for the binding sites of the EPS produced by *Synechocystis sp.* PCC 6803. This contrasts sharply with the synergistic interaction observed for *M. aeruginosa* PCC 7005.

## Influence of light intensity and cyanobacterial concentration on bloom formation

In the previous sections, we have identified biomass flocculation following a sudden increase in salinity and photosynthetic oxygen production as the physico-chemical mechanisms underlying the irreversible migration of the biomass. In natural conditions, the intensity of these processes is governed by the interplay between various environmental parameters. In order to determine whether the process of bloom formation described here could occur in a natural environment, we have assayed blooming for a range of cyanobacterial concentrations, light intensities and salt (BG11) concentrations on the cyanobacteria *M. aeruginosa* PCC 7005 and summarized the results in Fig. 4. For each set of parameters, we recorded whether bloom occurred and the time it took to appear. At a fixed cell density of $7 \cdot 10^6$ cell/ml (Fig. 4 A and D), decreasing the light intensity from 14 to 0.3 $\mu E.m^{-2}.s^{-1}$ had no influence on the salt concentration threshold (relative BG11 concentration of $\sim 4X$) necessary to induce a bloom. Below 0.3

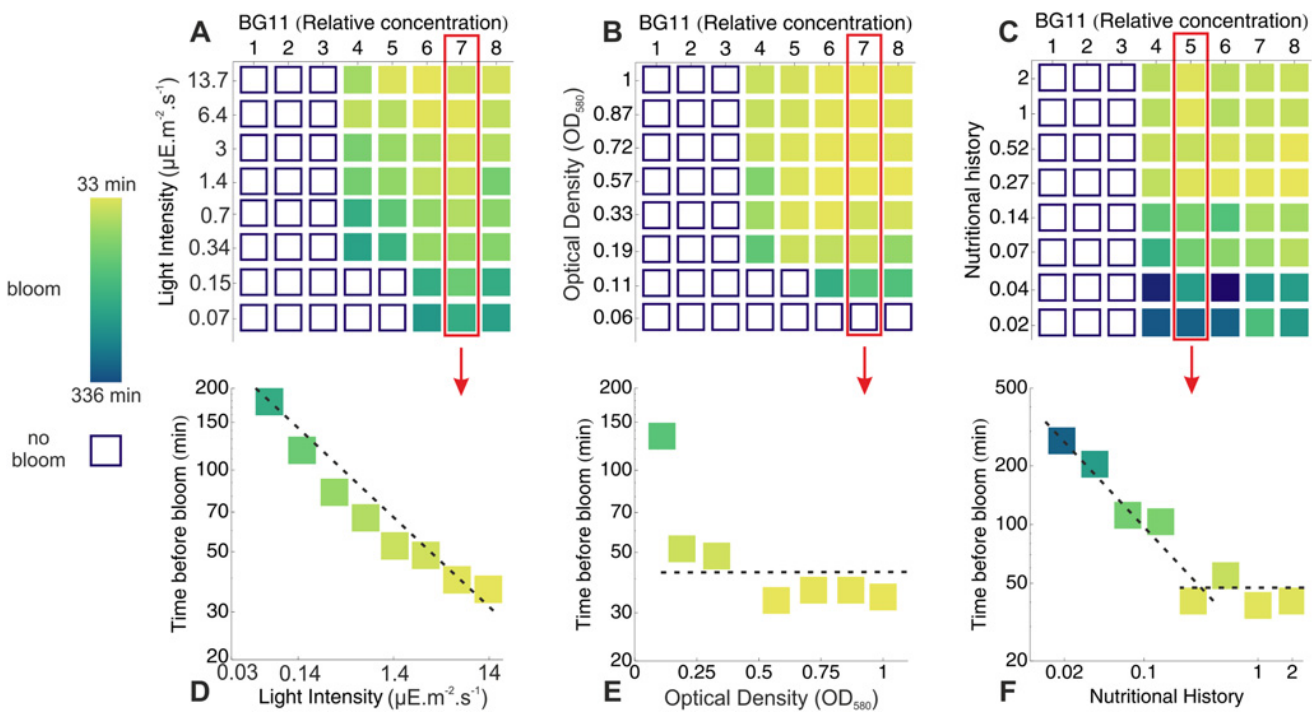

**Fig 4. Phase diagrams for bloom formation.** A-C: Phase diagrams for bloom formation. Open squares indicate that no bloom were observed within a time window of 24 hour while solid squares show that blooming occurred. The later are color coded with the timescale of bloom formation. For various values of the light intensity (A), initial cell density (B) and nutritional history (i.e relative nutrient concentration in the mother culture -C), 8 values of the salinity (relative BG11 concentration) were tested. D-F: Details of the phase diagrams A-C for fixed values of the salinity. Lowering the light intensity (D) increased the blooming timescale while changing the initial cell density $OD_{580}$ had little effect on this timescale (E). For nutritional history in the range 0.2 – 2X no effect on bloom formation were seen while below 0.2X the blooming timescale increased sharply (F).

doi:10.1371/journal.pone.0120906.g004

$\mu E.m^{-2}.s^{-1}$, this threshold increased slightly (relative BG11 concentration of $\sim 6X$). Decreasing the light intensity increased monotonously the timescale of bloom formation from $\sim 30$ to $\sim 200$ min. At a fixed light intensity of $\sim 7 \mu E.m^{-2}.s^{-1}$ (Fig. 4 B and E), changing the cell concentration in the range $3 \cdot 10^6 - 2 \cdot 10^7$ cell/ml had no effect either on the timescale or on the BG11 threshold of bloom formation. For cell densities of $2 \cdot 10^6$ cell/ml both the salinity threshold and the timescale of bloom formation rised sharply and no blooms were observed at cell densities of $10^6$ cell/ml and below. Similar results were also obtained by first inoculating the cells at low optical density with a high salt ([KCl] = 60 mM or [CaCl$_2$] = 20 mM) concentration and then allowing them to grow at $14 \mu E.m^{-2}.s^{-1}$. In that case, scum formation also occurred when the culture reached the critical cell concentration of $2 \cdot 10^6$ cell/ml.

## The timescale for bloom formation depends strongly on the nutritional history of the cyanobacteria

By comparing experimental data gathered on samples from exponential (OD$_{580}$ $\sim 1$—Fig 4-B) and late exponential to stationary phase (OD$_{580}$ $\gtrsim 2$—Fig 1-A), we noticed that samples from older cultures took a longer time to develop a bloom. While the age of a culture might not be an environmentally relevant parameter, this observation suggests that the nutritional history of a culture might influence the blooming process. In order to investigate this effect, we cultivated *M. aeruginosa* PCC 7005 in BG11 nutritive medium at various relative concentrations (0.02X to 2X) for three weeks and the cultures were then assayed for bloom formation by diluting the samples to an OD$_{580}$ = 0.35 and then adding concentrated nutrients to trigger bloom formation. Fig 4 C and F indicates that blooming occurred at the same salt concentration threshold (relative BG11 concentration of 4X), irrespective of the nutritional history. Similarly, no significant variations of the blooming timescale were observed when the nutrient concentration in the initial culture was decreased from 2X to $\sim 0.25X$. When this concentration was further decreased in the range 0.02-0.25X on the other hand, the timescale of bloom formation was strongly affected and increased from $\sim 40$ to $\sim 330$ minutes. Because blooming thresholds were not affected by the nutritional history, these data suggest that the increased timescale is not due to a change in EPS concentration but rather to a decreased O$_2$ production rate in cultures grown at lower nutrient concentration. In support of this hypothesis, we indeed measured a four-fold reduction in chlorophyll *a* content per cell in diluted medium (S3 Fig).

## Discussion

By modulating their buoyancy using internal gas vesicles, individual *M. aeruginosa* cells can migrate along the water column at speed up to 1 mm/h [21, 40, 41] while the collective migration mechanism studied here allows cells to migrate at speed $\gtrsim$1cm/h. When the salt concentration is above a salt and species-dependent threshold, cations screen the repulsive interactions between negatively charged EPS at the surface of cyanobacterial cells and cellular aggregates form. The concentrated photosynthetic activity in these clusters leads to O$_2$ supersaturation which nucleates into bubbles. In turn, trapped O$_2$ bubbles provide a lift force that drags most of the biomass at the surface of the water column where it accumulates in dense foamy layers strikingly similar to those observed in natural blooms. Moreover, this fast migratory process is irreversible as the foam remains stable for weeks. In this study we have focused on the detailed quantification of the mechanism leading to scum formation. The accurate determination of the thresholds in light intensity, cell concentration, salinity and chlorophyll content necessary to induce a massive migration was allowed by the use of purified laboratory cultures that can yield highly reproducible results. Although similar experiments performed on field samples are likely to remain more qualitative, it would nonetheless be interesting to test the mechanism identified in this study using natural samples. For example, environmental isolates of *M. aeruginosa* form large colonies and exhibit a complex vertical migration dynamics due to their dynamic glycan ballast that could compete with the oxygen-mediated upward migration. Although possibly interesting, these effects require an in-depth investigation and are therefore postponed to a future study.

The blooming process studied here occurs in a range of conditions that are environmentally relevant as similar and even higher values than the measured thresholds have already been measured in natural waterbodies for the salinity [42], light, nutrient concentration (for cultures grown in diluted BG-11 medium) [43, 44] and oxygen saturation levels [45]. Although the salinity thresholds for bloom in *M. aeruginosa* PCC 7005 and *Synechocystis sp.* PCC 6803 are low enough to be exceeded in natural environments [42], the competitive binding between monovalent and divalent cations in *Synechocystis sp.* PCC 6803 is likely to inhibit its ability to bloom in most waters. Indeed, while we successfully produced *Synechocystis* blooms in the laboratory, this specie is not usually associated with environmental blooms. By contrast, the synergistic binding of metallic cations with the EPS produced by *M. aeruginosa* PCC 7005 may explain the abundance of this specie in natural blooms. While cations also bind cooperatively in the toxic strain *M. aeruginosa* PCC 7806, this specie is less likely to bloom in nature as its salt thresholds are 3 to 6 times higher than in the nontoxic strain. Although both synergistic and competitive associations between cyanobacterial EPS and metals have been reported previously [36, 46, 47], the underlying mechanisms remain poorly understood and more work is needed to elucidate these binding processes. In this respect, similar experiments performed with cyanobacterial mutants of *Synechocystis sp.* PCC 6803 defective for EPS production [31] could provide valuable data to further understand the cell aggregation process.

Both the nutritional history and the light intensity were found to increase the timescale for bloom formation without any significant effect on the salinity thresholds. This variation can be explained by a decrease in the $O_2$ production rate per unit of biomass under low light intensities or in case of a weaker photosynthetic machinery in poor environments. Of course, other factors such as the temperature, can contribute to the efficiency of the photosynthetic $O_2$ production. While the timescale for bloom never exceeded $\sim$ 6 hours in our experiments, the $O_2$ accumulation rate in nature is likely to be smaller due to both convection and the respiratory activity of other species, two effects that are absent in our experiments. Because the timescale of bloom cannot exceed $\sim$ 12 hours (the maximum sunlight exposure), $O_2$ production may be the limiting factor for bloom formation and, while outside the scope of this study, it would also be interesting to conduct additional experiments with PSII-inactivated mutants defective in oxygen evolution.

Given the slow growth of cyanobacteria, blooming occurs without major increase of the overall biomass. We found that the initial cell density had little effect on either the salinity threshold or on the timescale of bloom. On the other hand, irreversible migration only occurred above a sharply defined concentration threshold of $\sim$ $10^6$ cell/ml ($OD_{580} = 0.05$). This ON/OFF behavior can be ascribed to the self reinforcing character of the blooming process. Indeed, as cyanobacteria migrate toward the free-surface, their effective concentration increases and the system moves away from the critical point for blooming. As such a concentration is already quite high in the environment, this suggests that other mechanisms must act to concentrate the micro-organisms before irreversible blooming can occur. In particular, temperature, light and nutrient gradients and interactions [48] have been shown to help localizing gas-vacuolate cyanobacteria near the surface of the water column, as mentioned in the introduction. While the mechanism described here is likely to occur essentially in the later phase of blooming, understanding surface scums formation is nonetheless of a considerable importance as these compact aggregates of cyanobacteria are in fact responsible for the most severe effects associated with blooms [49].

Beside the environmental relevance of the present study, exploiting the natural tendency to bloom of cyanobacterial cultures could have a valuable impact in an industrial context given the high potential of cyanobacteria for biofuel generation [50–52]. Indeed, the separation of microscopic organisms, whose density is close to that of water, from the culture medium in bioreactors is so far an expensive and challenging bio-engineering process, requiring the filtration or centrifugation of large volumes. Inducing the global agglomeration and migration of the biomass by raising the salinity could offer a fast, easy, inexpensive and environmentally friendly harvest solution in large pools, also because the recovery and recycling of salty nutrients is easy.

## Supporting Information

**S1 Fig. Evolution of the blooming process for various BG11 concentration.** Assay for bloom formation in *M. aeruginosa* PCC7005. Aliquots from a culture in the late exponential phase (OD $\sim$ 2) were resuspended to a final optical density of 0.35 in BG11 medium at various relative concentrations (1X being the reference given in the Material and Methods section) indicated on top of each tube. Series of tubes are shown at three different times: 0, 150 and 300 min. after nutrient pouring. Note that for relative nutrient concentration above $\sim 20X$, there is a distinctive yellowing of the biomass while the lower fraction progressively becomes black. These changes in coloration can be attributed to cell lysis together with the release of the cyanobacterial pigment phycocyanin.
(EPS)

**S2 Fig. Effect on sonication on blooming.** Series of pictures showing that sonication of cyanobacterial culture collapsed the gas vesicles but did not prevent blooming. Assay for bloom formation in *M. aeruginosa* PCC7005 following sonication. Aliquots from a culture in the late exponential phase (OD $\sim$ 2) were resuspended to a final optical density of 0.35 in BG11 medium 1X (negative control) and 4X (positive control) to induce blooming. Other aliquots were first sonicated and then resuspended at the same BG11 and cell concentration. In the sonicated sample at a BG11 concentration of 1X, there is a clear sedimentation of the cyanobacterial cell, indicating that the gas vesicles were indeed collapsed by sonication. This did not prevent the sonicated sample from blooming.
(EPS)

**S3 Fig. Chlorophyll a content.** Cultures of *Microcystis aeruginosa* PCC 7005 were grown at different BG11 concentrations (relative concentration between 0.01X and 1X) for three weeks under a 14h/10h light cycle (6 $\mu$E.m$^{-2}$.s$^{-1}$) at 25°C. The sample were then tested for bloom formation and their chlorophyll a content was measured. 1ml aliquots of the cultures were centrifuged at 13000g for 1min. The pellets were then resuspended in 90% methanol and allowed to sit 5 min in the dark to extract the chlorophyll. The samples were then centrifuged again and the optical density OD$_{665}$ of the supernatant was measured at 665nm. The optical density was converted in chlorophyll a concentration through the formula: chl a ($\mu$ g/ml) = 12.7OD$_{665}$.
(EPS)

**S1 Movie. Movie showing the evolution of the blooming process for various BG11 concentration.**
(AVI)

## Acknowledgments

We thank Alexandre Gelabert for lending us the oxygen probe, Maxime Costalonga for his help with the Navitar objective, Laurent Royon and Imane Boucema for discussions about the rheology measurements, Florent Carn for assistance with the zetasizer and Valérie Gautier for her help with the EPS extraction.

## Author Contributions